# Inferring Material Parameters from Current-Voltage Curves in Organic Solar Cells via Neural-Network-Based Surrogate Models

*Eunchi Kim\*, Paula Hartnagel, Barbara Urbano, Leonard Christen, Thomas Kirchartz\**

**Affiliations**
E. Kim, P. Hartnagel, B. Urbano, L. Christen, T. Kirchartz
IMD3-Photovoltaik
Forschungszentrum Jülich
52425 Jülich, Germany
E-mail: e.kim@fz-juelich.de
E-mail: t.kirchartz@fz-juelich.de
T. Kirchartz
Faculty of Engineering and CENIDE
University of Duisburg-Essen
Carl-Benz-Str. 199, 47057 Duisburg, Germany

## Abstract

Machine learning has emerged as a promising approach for estimating material parameters in solar cells. Traditional methods for parameter extraction often rely on time-consuming numerical simulations that fail to capture the full complexity of the parameter space and discard valuable information from suboptimal simulations. In this study, we introduce a novel workflow for parameter estimation in organic solar cells based on a combination of numerical simulations and neural networks. The workflow begins with the selection of an appropriate experimental dataset, followed by the definition of a device model that accurately describes the experiment. To reduce computational complexity, the number of variable parameters is carefully selected, and reasonable ranges are set for each parameter. Instead of directly fitting the experimental data using a numerical model, a neural network was trained on a large dataset of simulated results, allowing for efficient exploration of the high-dimensional parameter space. This approach not only accelerates the parameter estimation process but also provides valuable insights into the likelihood and uncertainty of the estimated parameters. We demonstrate the effectiveness of this method on organic solar cells based on the PBDB-TF-T1:BTP-4F-12 material system, demonstrating the potential of machine learning for rapid and comprehensive characterization of emerging photovoltaic materials.

## 1. Introduction

The speed and simplicity of solution processing combined with the huge chemical space of molecular semiconductors[1-3] has led to rapid improvements in efficiency of organic solar cells over the last years, with some recent reports exceeding the 20% efficiency threshold.[4-7] Despite significant advantages of organic photovoltaics such as low material consumption, abundance and non-toxicity of the involved elements, low weight, flexibility and the potential for fine tuning of transparency, color and absorption threshold, even surpassing the 20% milestone, still leaves a significant gap to other photovoltaic technologies such as silicon and halide perovskites.[8] These additional losses in organic solar cells are a combination of optical losses, charge transport losses[9] and recombination losses at open circuit (see e.g. Figure 1 in ref. [10] and explanations in ref. [11]). Reducing these losses requires the ability to quantify the properties of materials and interfaces and relate them to changes in device performance and eventually also stability. While the material properties of classical solar cell materials, such as crystalline silicon, are widely known and tabulated,[12, 13] comparable data for emerging solar cell material systems, such as molecular semiconductors or halide perovskites, are scarce. This lack of knowledge about material properties can on the one hand be attributed to the high variety of organic semiconductors but is also caused by the strong correlation between material parameters and experimental observables. Whereas optical measurements directly yield the absorption properties of the material, most measurements used to quantify the electronic properties depend on a multitude of these electronic parameters at the same time.[14] In these cases, simple analytical models for the extraction of these parameters are not sufficient; however, numerical simulations are required to understand the measured data. The problem with numerical device simulations is that they allow us to simulate an experiment based on known material properties (mobility, recombination coefficient, band gap, etc.); however, they do not allow us to invert the problem directly. Therefore, one cannot give the simulation tool a current-voltage ($JV$) curve and ask for the material properties used to create such a $JV$ curve. Furthermore, particularly in the case of a single $JV$ curve, many possible combinations of material properties would have led to the same curve within a small margin of error. This thought experiment immediately suggests two challenges that must be overcome: (i) a time-efficient inversion of numerical device simulation models and (ii) a probabilistic rather than deterministic approach to parameter inference. Challenge (i) has, in the past, primarily been approached using fitting models, where a drift-diffusion model was executed with variable material parameters as input until the output of the model best matched the experiment.[15-18] This approach allowed the quantification of material parameters but did not quantify likelihoods, did not notice how many different sets of parameters reproduced the data, and was not time efficient, as the simulation would have to be run thousands of times to reach a good fit.

One attempt to address these disadvantages is to implement machine learning in the process of estimating material parameters using simulation tools. Attempts have been made to train a neural network with simulated data to predict the material parameters based on experimental input data.[19, 20] This approach yields a parameter combination that fits the experimental data. However, information on the surrounding parameter space that provides confidence in the fit remains unknown. To retrieve this information, Bayesian inference has been used, in which the parameter space is sampled statistically based on Bayes' theorem[21] instead of using an optimization algorithm. Although Bayesian parameter estimation was used by several different groups,[22-25] the computational costs remained immense.[24] To reduce computation time, Ren et al. replaced the simulation software with a neural network that was trained with simulated data on gallium arsenide solar cells[26], and the method was later applied to thin-film photovoltaics[27]. Therefore, the Bayesian inference process was sped up significantly. This approach of using a neural network instead of simulation software has not yet been applied to organic photovoltaics. While the field of machine learning has also found growing interest,[28] efforts have mainly focused on material screening,[29-33] optimizing the donor-acceptor ratio [34, 35] or for device fabrication.[36-38] However, in the field of device characterization, the opportunities arising from machine learning are relatively unexplored in organic photovoltaics with few notable exceptions by Majeed et al.[19], Raba et al.[23], and Hußner et al.[39]. However, the combination of fast computation with a neural network and the exploration of a high-dimensional parameter space has not been investigated so far. Such an approach goes beyond the traditional fitting procedure, where all simulations, except the best fit, are discarded and their information is lost; instead, all simulations with the numerical solver are used for neural network training. Because the neural network not only stores this information but can also interpolate between points, new opportunities for the rapid analysis of a material system arise. Therefore, we introduce a parameter estimation workflow using a neural-network-based surrogate model and apply the workflow to the problem of parameter inference based on current-voltage curves of organic solar cells based on the active layer blend PBDB-TF-T1:BTP-4F-12[40].

## 2. Methods

**Figure 1** illustrates the workflow, starting with the choice of experiment (**Figure 1a**). Applied to the context of organic solar cells, this experiment could be a simple *JV* curve or a capacitance-voltage curve, but also a combination of several different measurements because they could contain different information on the material parameters. After choosing the experiment, a device model must be defined (**Figure 1b**) that accurately describes the experiment performed. Such a device model can range from an analytical equation[41-45] to software, such as ASA[46-49] or SCAPS[50-53]. Here, we use the drift-diffusion solver ASA, which combines the speed of execution with an easy method to interface with MATLAB or Python. However, such device simulations require many input parameters. For example, a drift-diffusion simulation using an optical solver requires 40 parameters to cover the most common recombination mechanisms. Because every parameter adds a dimension to the investigated parameter space, the number of variables must be reduced (**Figure 1c**) to maintain the computational effort at an acceptable level. Here, on the one hand, it is crucial to choose the parameters that have the biggest influence on the system but also to set the remaining parameters to a reasonable value. In addition, setting a reasonable range for the variables is important as it should be ensured that the expected values lie within that range. Once the variable material parameters and their boundaries were selected in a traditional fitting routine, a fitting algorithm was applied to run the model, that is, the numerical simulations, until it reached an optimum. This procedure is time-consuming, especially considering that the information gain is limited to the best fit, whereas all other simulations during fitting are discarded. Therefore, instead of directly using the model to fit the experimental data, a training dataset was generated (**Figure 1d**). The goal is to create a large set of parameter combinations and their corresponding outcomes, such as *JV* curves, that cover the entire parameter space. Many combinations of material parameters and simulation results are required to train the neural network (**Figure 1e**). The more material parameters are selected to be a variable, the more input data must be used for training the – then higher dimensional – neural network. Because the neural network can interpolate between these data points in a fraction of the time required for numerical simulations, it can replace the device simulator. The goal is for the neural network to accurately predict information with any given parameter set that lies within the boundaries defined by the training data simulations. Having a neural network that can replace the device model enables high-throughput predictions that can be compared with experimental data. By doing so, one can fit the experimental data for the parameter set that best describes the data (**Figure 1f**), which would otherwise be challenging in high-dimensional space. However, going beyond this best point is also possible by scanning the entire parameter space and calculating the error of the predicted data at each point compared to the actual experiment (**Figure 1g**). The error evaluation enables the estimation of the posterior probability density function based on Bayes' theorem.[22] Consequently, it enables the quantification of the dependence of the probability density on each material parameter within the parameter space. Therefore, creating a framework, as shown in the upper panel of **Figure 1**, enables high-throughput analysis of the experimental data.

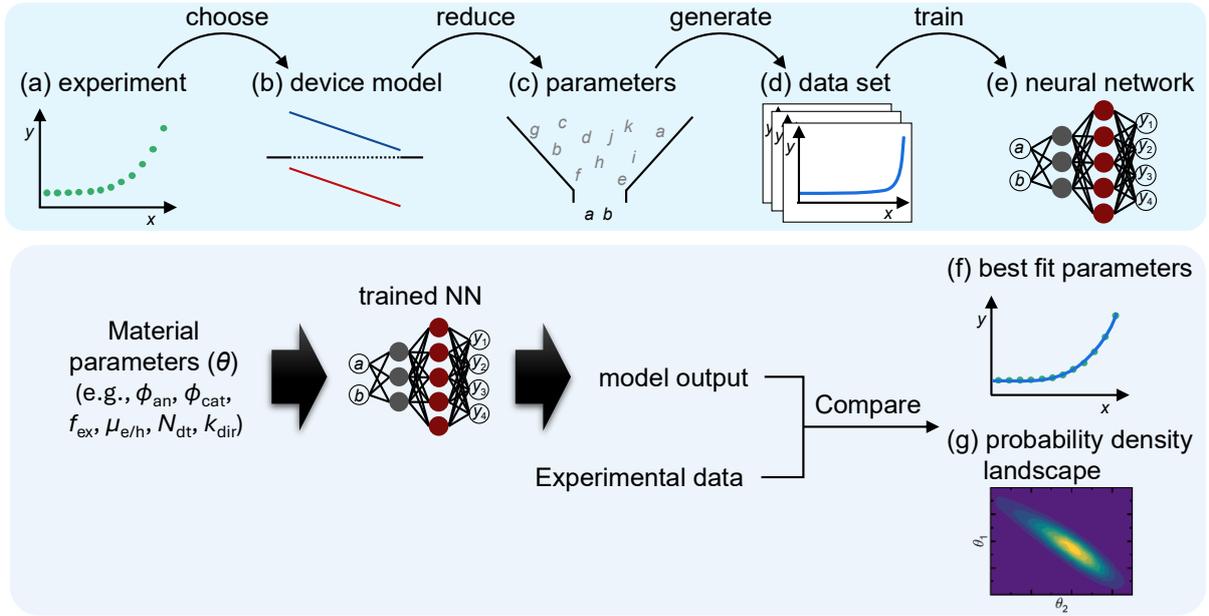

**Figure 1.** Schematic of the workflow. After collecting (a) experimental data, (b) a device model that best describes the experimental data must be chosen. Often, a reduction in the dimensionality of the problem is required, meaning that (c) variable material parameters are selected, whereas others are fixed. Using the device model, (d) a training dataset was generated by varying the selected parameters. (e) A neural network (NN) was then trained with the data and serves as a forward device model. Comparing the output of the neural network to the experimental data allows the calculation of an error and likelihood, thereby (f) finding the best parameter set with a fitting procedure and (g) visualizing the probability density as a function of material parameters. Establishing an NN model through the first five steps (a-e) facilitates high-throughput analysis of a solar cell.

## 3. Results

To apply the parameter estimation routine presented above, we chose current-density voltage curves, a readily available characterization tool. Because material properties interdependently influence the shape of the $JV$ curves, it is difficult to robustly narrow down the parameters with one type of experiment. Therefore, we decided to use a set of thickness-dependent or light-intensity-dependent $JV$ characteristics.

With the knowledge of what experimental data should be analyzed, in the next step, the number of variable device parameters for the simulations and later fitting must be reduced. Here, we decided to allow the algorithm to vary the interfacial energy level alignment of the device, represented by the injection barriers for the anode $\phi_{an}$ and cathode $\phi_{cat}$. For organic solar cells, where exciton dissociation is a crucial intermediate step during charge generation, we included the exciton dissociation probability $f_{ex}$ as another parameter to be varied. For recombination, both direct recombination and trap-assisted recombination via the tail states and deep defects are considered. Among these, only two were treated as variables: the direct recombination coefficient $k_{dir}$ and the density of defect states in the center of the band gap $N_{dt}$. When defining the allowed charge states of deep defect states, it is important to consider where they would create a space-charge region. Acceptor-like defects create a region with a high electric field close to the cathode, whereas donor-like defects create a space-charge region near the anode. Because most charge carriers are generated close to the transparent anode, the space charge in this region promotes charge extraction and causes the thickness dependence of the short-circuit current density to exhibit interference maxima. When the space-charge region is opposite to the illuminated contact, the system is limited by the diffusion through the low-field region and the interference maxima in the $J_{sc}$-thickness relation are less pronounced.[54] In this work, donor-like defect states were included to enable space-charge formation at the illuminated anode. Lastly, we varied charge-carrier mobilities, while both electron and hole mobilities are fixed to be equal. Six variable material parameters were selected as listed in **Table 1**.

Table 1. Material parameters chosen as variable parameters with upper and lower boundaries for training data generation.

| Parameter | Lower Boundary | Upper Boundary |
| --- | --- | --- |
| Injection barrier anode $\phi_{an}$ (eV) | 0 | 0.4 |
| Injection barrier cathode $\phi_{cat}$ (eV) | 0 | 0.4 |
| Exciton dissociation probability $f_{ex}$ | 0.6 | 1 |
| Electron/hole mobility $\mu$ (cm$^2$V$^{-1}$s$^{-1}$) | $10^{-4}$ | 1 |
| Density of deep trap states $N_{dt}$ (cm$^{-3}$) | $10^{14}$ | $10^{17}$ |
| Direct recombination coefficient $k_{dir}$ (cm$^3$s$^{-1}$) | $10^{-12}$ | $10^{-6}$ |

All remaining parameters required in the simulations were kept constant. After determining the ranges in which the material parameters are varied, the training data can be simulated using ASA. For this purpose, we generate a set of parameter combinations as a Sobol sequence, which ensures good coverage of the entire parameter space.[55] For this work, we simulated 300,000 *JV* curves for each light intensity. Depending on the number of points simulated in the active layer with ASA, that is, the active layer thickness, the training data generation can take between 12 and 22 h. Once the simulation of the *JV* curves was completed, the simulated datasets were used for training of the neural network. Further details on the training process can be found in Supporting Information.

### 3.1. Test on Synthetic Data with Known Parameters

With the neural networks at hand that can replace ASA and speed up computational work, we can proceed to explore the entire parameter space and compare the predictions of the neural network ($J_{NN}$) to the experimental data ($J_{exp}$). However, high dimensionality of the parameter space makes it challenging to interpret and visualize the multivariate probability distribution. To address this, we first demonstrate our approach using synthetic data with known material properties, where the criteria for success are well-defined. Firstly, we employ a genetic algorithm to rapidly scan the multi-dimensional parameter space and identify the optimum point where there is the least difference between the experimental data and the output of NN model. Subsequently, we compute posterior probability distribution of either around this optimum point or over the parameter space of interest. This enables us to verify whether the fitting algorithm has found the global optimum and to observe how the uncertainties associated with the inferred parameters evolves as new experimental data is included. Furthermore, visualizing these distributions in one- or two-dimensional plots provides a glimpse of how the parameter space looks like.

To fit the experimental data, we used a covariance matrix adaptation evolution strategy (CMA-ES) algorithm[56], one type of an optimization algorithm for solving high-dimensional problems with interdependent variables.[27, 57, 58] The algorithm searches for the optimal point according to a manually defined evaluation function, which we specified as root-mean-square error between $J_{NN}$ and $J_{exp}$ over the voltage sweep normalized by the measurement uncertainty. (**Equation S1**) The uncertainty of a measurement represents the variance or error caused by characterization tools such as light or voltage source. In this work, we only included the variance of the light source and assumed that the instability of the source-measure unit (SMU, Keithley 2450) is negligible.

Subsequently, we fitted a simulated dataset and compared the fitting results with actual input values. **Figure 2** visualizes the fitting algorithm where only a subset of the six material parameters were shown, focusing on the transport and recombination parameters, namely the charge carrier mobility $\mu$, defect density $N_{dt}$, and direct recombination coefficient $k_{dir}$. The points in the figure correspond to candidate combinations of six parameters, and the color mapping indicates the sequence index of those possible solutions suggested by the algorithm. In other words, the color mapping from dark green to yellow reflects the progression of the search over time. Initially, the algorithm scans the entire parameter space and then gradually converges towards the optimum point. Therefore, despite the high dimensionality of the problem, the fitting algorithm appears to find the optimum value that is close to the actual parameters. Still, to avoid converging to local minimum, we typically perform 7~10 cycles of this process and select the global minimum as the best-fit parameters $\theta_{\text{best-fit}}$. A single run of the algorithm requires between 10000 and 30000 calculations, depending on the initial setting. Given that the neural network can produce >3000 $JV$ curves per second, the entire fitting procedure is completed in a few minutes with a conventional laptop with a NVIDIA T550 graphics card. It is a significant reduction in computation time compared to the numerical solver ASA, which would take up to eight hours for a single fitting run.

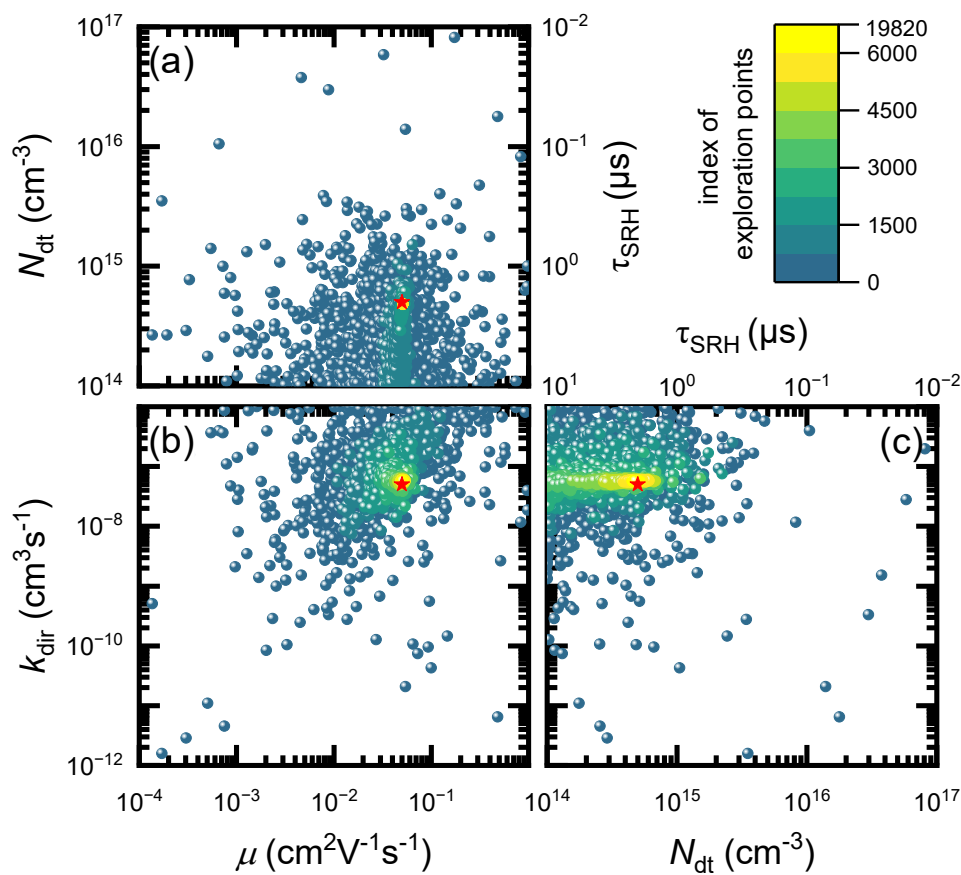

**Figure 2.** Visualization of the CMA-ES fitting algorithm on synthetic data, where the actual material parameters (red star) are known. The color bar indicates the sequence index of the exploration points. As the algorithm navigates the six-dimensional space to find the global optimum point with regard to the predefined evaluation function, it converges to the true values within 6000 points, completing the entire process in approximately 20s on a conventional laptop with a graphic processing unit NVIDIA T550. This efficiency is enabled by a neural network model that delivers high-speed computation, generating >3000 $JV$ curves per second.

We expect that using an optimization algorithm to find the best-fit values would be particularly well-suited for high-throughput analysis of solar cells, facilitating the speed of the workflow. Moreover, replacing a numerical solver with a NN model offers an additional advantage: it enables the estimation of probabilistic values for parameter combinations across the entire parameter space and tracking of changes in the probability distribution following additional feeding of information. By applying Bayes' theorem[21]

$$P(\theta|J_{\text{obs}}) = \frac{P(\theta)P(J_{\text{obs}}|\theta)}{P(J_{\text{obs}})} = \frac{P(\theta)P(J_{\text{obs}}|\theta)}{\int P(\theta)P(J_{\text{obs}}|\theta)d\theta}, \qquad (1)$$

we can estimate the posterior probability $P(\theta|J_{\text{obs}})$, which represents the probability that a specific set of parameters corresponds to the actual material parameters given a particular observation or dataset. Here, $P(\theta)$ is the prior probability, $P(J_{\text{obs}}|\theta)$ is the likelihood or the probability distribution of the observed data $J_{\text{obs}}$ at the given set of parameters $\theta$, and $P(J_{\text{obs}})$ is the marginal likelihood or normalization factor to ensure the integral of the posterior distribution over the parameter space to be one. **Figure 3a** illustrates the workflow of Bayesian parameter estimation. The process begins with defining a prior probability distribution, assumed to be a continuous uniform distribution, representing our initial state of knowledge when no data is available. Next, we calculate the likelihood from the error between the model output ($J_{\text{NN}}$) and the experimental data ($J_{\text{exp}}$), incorporated with the uncertainty associated with the measurement apparatus (see **Equation S1**). The denominator of **Equation 1**, representing the marginal likelihood, is approximated using a Monte-Carlo method[59], which involves randomly sampling the entire parameter space (see Part C in Supporting Information for detailed explanation). The resulting posterior probability density function obtained through this process is iteratively updated as new observations are available.

To demonstrate the Bayesian framework, we generated four distinct light-intensity-dependent $JV$ characteristics with a numerical solver (**Figure 3b**) and computed the posterior probability distribution for each observation of a $JV$ curve. The Bayesian framework allows us to systematically quantify how our belief about the material parameters is shaped as new data becomes available, given the context of previous observations. **Figure 3c-h** illustrates a multivariate posterior probability distribution projected onto discrete points in a one-dimensional space, obtained by integrating over the remaining five dimensions. The color gradient, ranging from yellow to dark purple, indicates the sequential update of the posterior probability, as the number of observed $JV$ curves $n_{\text{JV}}$ increases. For some parameters (**Figure 3e-f and 3h**), the change in the probability distribution is evident from the initial dataset, and subsequent additions of data strengthen this trend, with the peak position remaining relatively consistent and the peak narrowing as more information is updated. However, the projected posterior as a function of injection barriers (**Figure 3c-d**) remains relatively unchanged from the uniform prior distribution. In contrast, the posterior distribution of $N_{\text{dt}}$ undergoes a significant change, where the initial peak observed in the first sequences becomes less significant as additional information is updated, and the highest probability density point shifts to the left. This suggests that a small number of observations may lead to inaccurate parameter inference.

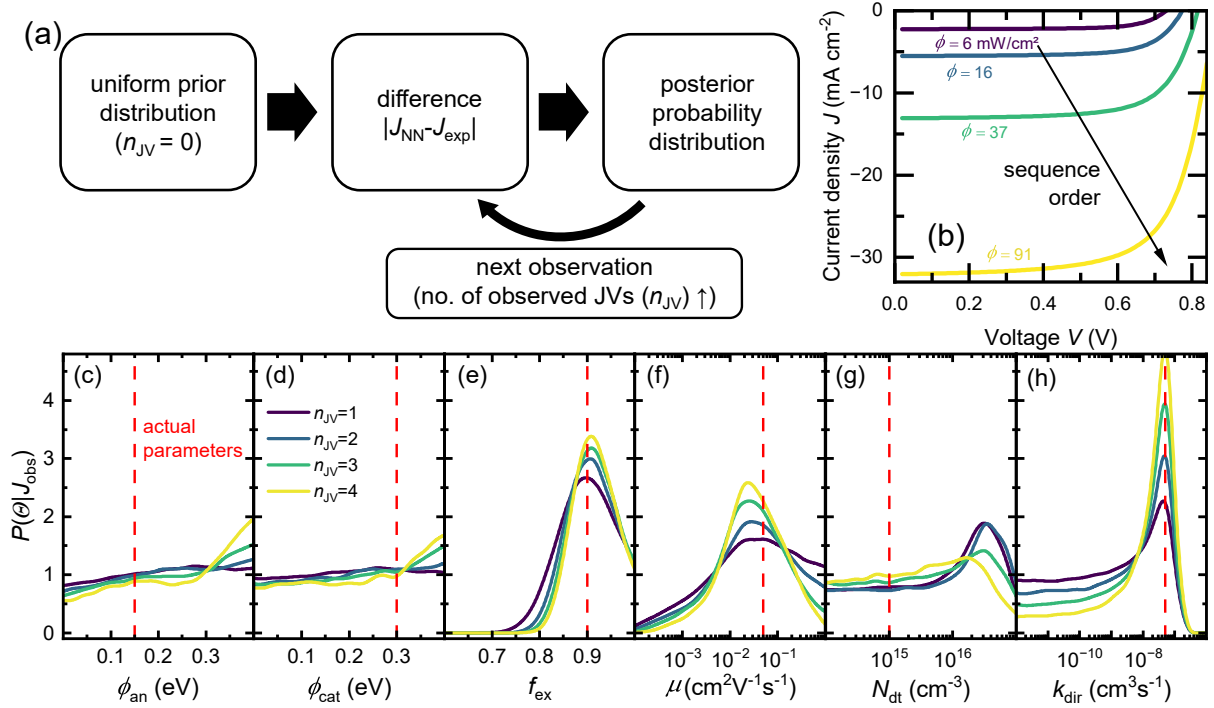

**Figure 3.** (a) Schematic description of the Bayesian inference framework. (b) Current-voltage characteristics of a synthetic dataset where light intensities were varied. The inclusion of a new *JV* curve begins with the lowest light-intensity curve and ends with the highest one. (c-h) Projection of the multivariate posterior probability distribution $P(\theta|J_{obs})$ onto a one-dimensional parameter space for the synthetic data shown in panel (b). The red dashed line represents the true parameter values. Color coding indicates the number of light-intensity-dependent *JV* characteristics $n_{JV}$ used to estimate posterior probabilities. As new information is incorporated into the prior belief, the posterior probability is updated, either reinforcing the prior belief or altering the shape of the posterior density function.

Additionally, maxima of projected probability distributions at $n_{JV} = 4$ for $\phi_{an/cat}$, $\mu$, and $N_{dt}$ do not coincide with the true parameter values. Here, it is important to note that a high integral value can arise from various scenarios, such as a small volume with high probabilities or a large volume with moderate-high probabilities. Unfortunately, it can be challenging to distinguish between these cases simply by examining the projected probability due to the high dimensionality of the parameter space. To address this issue, we can further analyze the posterior probability densities of material parameters around the best-fit values, where high values of probabilities are likely to be located. Subsequently, we plot the posterior probability densities where one parameter is varied while the others are fixed to the $\theta_{\text{best-fit}}$ found by CMA-ES, as illustrated in the green line plots on the diagonal of **Figure 4**. Similarly, we examine the 2-dimensional posterior probability distributions where two parameters are varied while the rest is fixed to the $\theta_{\text{best-fit}}$. The contour plots in **Figure 4** represent these distributions. Unlike projected probability distributions, the probability distributions sliced through the point $\theta_{\text{best-fit}}$ have their maximum peak approximately at the actual parameter value, verifying the accuracy of the fitting algorithm. Given the significant impact of the point $\theta_{\text{best-fit}}$ on the projected probabilities, summing up the relevant contour plots as a function of one parameter of interest can provide a meaningful insight into interpreting the projected probability distributions, although it does not possess the same mathematical properties as a probability density function derived through integration. For example, the maximum peak of the blue line plot for $\mu$ is shifted lower than the actual value. This discrepancy might be attributed to the large volume of moderate-high probabilities observed in the 2D plots with injection barriers or with $k_{dir}$.

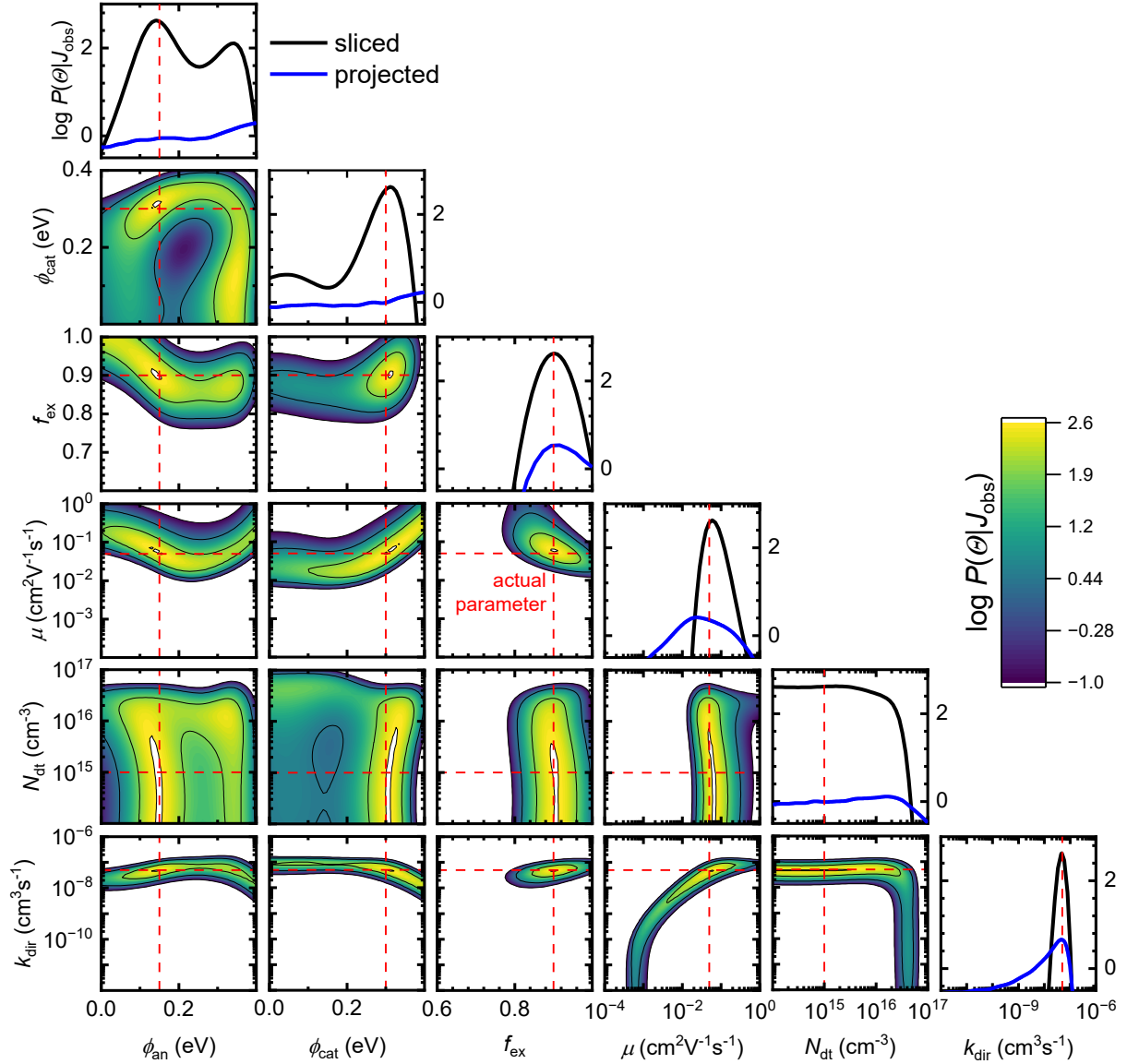

**Figure 4**. Corner plot of the posterior probability distributions for a synthetic dataset where the actual parameters are presented in red dashed line with the four illumination-dependent current-voltage curves included. The black line plots on the diagonal represent a one-dimensional slice through the six-dimensional parameter space at best-fit values obtained from the genetic algorithm CMA-ES, whereas the contour plots illustrate a two-dimensional slice. The blue line plots on the diagonal show a projected probability onto one-dimension. Maxima of sliced one- and two-dimensional plots coincide well with the actual parameters, highlighting the accuracy of the optimization algorithm. On the other hand, maximum peaks of the projected probabilities partially deviate from the true values. The deviation should be cautiously interpreted since the projected probabilities can arise from various scenarios: small volume of high probabilities or large volume of moderate-high probabilities. In this regard, the 2D contour plots of probability distribution around the $\theta_{\text{best-fit}}$ could provide another aspect in understanding multi-dimensional parameter space. For example, the shift of maximum peak for $\mu$ might be attributed to large volume of moderate-high probabilities observed in the 2D plots with injection barriers or with $k_{\text{dir}}$.

Furthermore, we investigate how each characterization data influence our belief in material parameters by calculating Kullback–Leibler divergence $D_{\text{KL}}$.[60] This metric compares the prior and posterior probability distributions to show how much our knowledge about the parameters improves after including the data.[61, 62] In other words, it quantifies how much information for material parameters is gained when updating from the prior to the posterior distribution. Kullback-Leibler divergence is defined via

$$D_{\text{KL}} = \int P(\theta) \ln \frac{P(\theta)}{Q(\theta)} d\theta, \qquad (2)$$

where $P(\theta)$ is the posterior probability distribution and $Q(\theta)$ is the prior.

**Figure 5** illustrates the Kullback–Leibler divergence of the six material parameters evolving with the observation of a new data. The probability distributions in **Figure 3c-h** are used to estimate the entropy using **Equation 2**, where the $Q(\theta)$ is set as the uniform prior. Another version of relative entropy where the $Q(\theta)$ is defined as the immediate prior probability distribution can be found in SI (i.e. for observation 1, the prior is uniform, for observation 2, the prior is the posterior of observation 1, etc…). For $\phi_{\text{an}}$ and $\phi_{\text{cat}}$, the $D_{\text{KL}}$ remains approximately zero, indicating that the probability distribution has not changed significantly from the initial uniform prior distribution. This is mathematical evidence for the fact that just measuring $JV$ curves is insufficient to quantify the injection barriers. This observation is supported by the results shown in **Figure 3c-d**. In contrast, the $D_{\text{KL}}$ of $f_{\text{ex}}$, $\mu$, and $k_{\text{dir}}$ is high after the update with the lowest light-intensity $JV$ data. The $D_{\text{KL}}$ of $f_{\text{ex}}$ saturates upon further information updates, while that of $\mu$ and $k_{\text{dir}}$ continuously increases. The saturation of the information entropy relative to the uniform prior suggests that the subsequent light-intensity-dependent $JV$ curves do not provide sufficiently useful information to further quantify $f_{\text{ex}}$. This is understandable, as in our model, $f_{\text{ex}}$ is just a voltage- and light-intensity-independent factor that reduces the photocurrent relative to the result from purely optical simulations. In contrast, mobility and recombination coefficient have an influence on fill factor that depends on light intensity[54, 63, 64] and, therefore, confidence in their inferred values benefits from including data taken at different light intensities. For $N_{\text{dt}}$, the $D_{\text{KL}}$ increases at the beginning and then goes back to zero, indicating that the posterior distribution becomes similar to a uniform distribution.

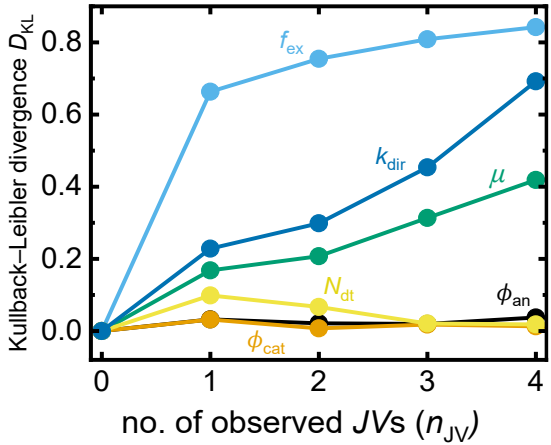

**Figure 5.** Kullback-Leibler divergence $D_{\text{KL}}$ of material properties compared to the uniform prior distribution for a synthetic dataset. This method measures the amount of information gained from $JV$ curves. A substantial amount of information is gained after observation of four $JV$ curves for $f_{\text{ex}}$, $\mu$, and $k_{\text{dir}}$, while it is challenging to determine injection barriers and defect densities.

The information gained during parameter inference can vary across the six material properties, depending on the specific $JV$ curve. To obtain a general overview of information extracted from four illumination-dependent $JV$ curves, we generated simulated datasets with randomly varied material parameter combinations and calculated the $D_{\text{KL}}$, as shown in **Figure 6a**. Consistent with previous observations, inferring $f_{\text{ex}}$, $\mu$, and $k_{\text{dir}}$ from these datasets is generally less challenging. Notably, the $D_{\text{KL}}$ values for $k_{\text{dir}}$ exhibit significant variation across datasets, reflected in a high standard deviation. This may result from the dominance of one recombination mechanism masking the characteristics of the other. To verify this, we plot $D_{KL}$ for $N_{dt}$ and $k_{dir}$ as a function of the ratio of direct to total recombination current $J_{rec,dir}/J_{rec,total}$ at short circuit, as illustrated in **Figure 6b**. When direct recombination dominates over recombination via defect states, it is likely to obtain substantial amount of information for inferring $k_{dir}$, while the posterior probability distribution for $N_{dt}$ remains broad and less informative.

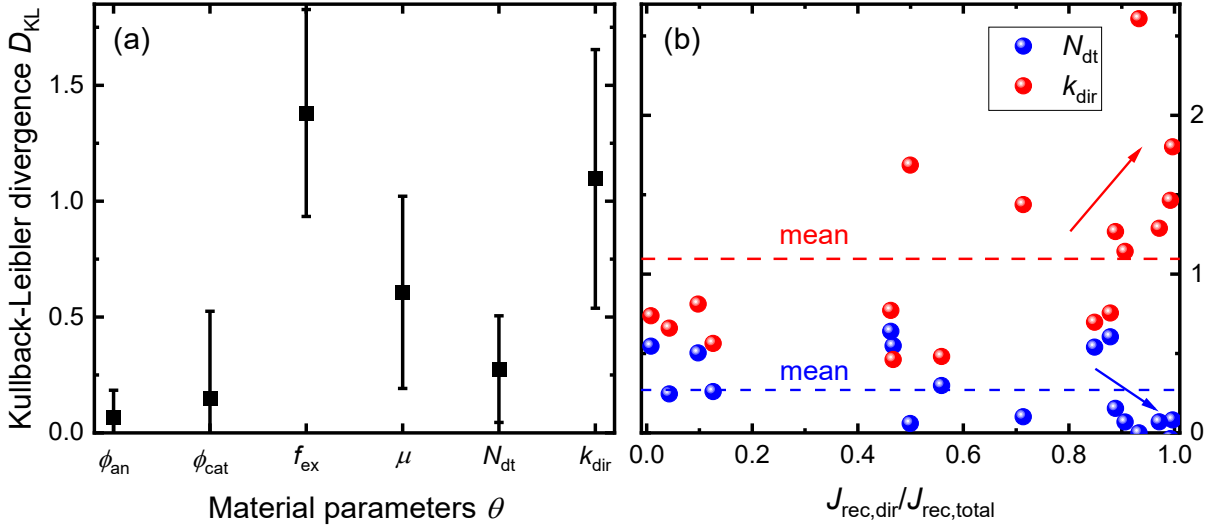

**Figure 6.** (a) Mean and standard deviation of Kullback-Leibler divergence $D_{KL}$ of material properties obtained from simulated datasets with randomly varied material parameter combinations. Each dataset consists of four light-intensity-dependent $JV$ curves. (b) $D_{KL}$ for defect density $N_{dt}$ (blue) and direct recombination coefficient $k_{dir}$ (red) plotted against the ratio of direct to total recombination current $J_{rec,dir}/J_{rec,total}$ under short-circuit condition. The dashed line represents the mean value of $D_{KL}$ for all data points in the plot.

### 3.2. Bayesian Parameter Estimation on Organic Solar Cells

After validating our approach by testing it on synthetic data, now we proceed to the analysis of experimental datasets that we obtained from the lab. We began by measuring thickness-dependent and light-intensity-dependent $JV$ characteristics of organic solar cells based on a binary system PBDB-TF-T1:BTP-4F-12. A detailed description of the solar cell fabrication procedure can be found in the Supporting Information. The results of experimental analysis are presented as points in **Figure 7**, where the power conversion efficiencies are plotted against thickness for four discrete light intensities: 6, 16, 37, and 91 mW/cm². The plots reveal that the highest efficiencies are achieved at an active layer thickness of approximately 70 nm, with a secondary maximum at around 230 nm. **Figure S11**, displaying the short-circuit current densities $J_{sc}$, open-circuit voltage $V_{oc}$, and fill factor $FF$ as a function of thickness, reveals that the efficiency of thicker cells is limited primarily by $FF$. The gain in $J_{sc}$ due to increased absorption of light is offset by reduced charge collection efficiency.

Similar to the case of PBDB-TF-T1:BTP-4F-12, performance of most organic solar cells for high thickness (> 200 nm) is limited by the charge collection efficiency. Indeed, this imperfect charge carrier transport is extended to thin cells due to inherent nature of organic semiconductors – low mobilities. Previously, several attempts were made to express the charge collection efficiency analytically to understand losses from non-radiative recombination, while using either illumination- or thickness-dependent data to infer material properties of the system.[9, 45, 48, 54, 65, 66] Accordingly, we employ the same workflow of utilizing illumination- and thickness-dependent data, which are expected to provide insight into recombination dynamics and, thus, $JV$ characteristics. In addition to extracting material parameters from $JV$ curves, we aim to investigate how far these inferred parameters can be used to predict device performance at different thicknesses or light intensities which were not used in the inference.

Accordingly, we analyzed four different sets of current-voltages curves, which are highlighted as red points in **Figure 7**. Two of them are comprised of light-intensity-dependent $JV$ curves, while the thickness is fixed. Specifically, the "thin" set corresponds to a fixed thickness of 83 nm, while the "thick" set has a fixed thickness of 247 nm. The other two sets consist of thickness-dependent $JV$ curves, where light intensities are varied. One of these sets, denoted as "dim", is measured at the lowest light intensity of 6 mW/cm², and the fourth set is referred to as "bright", representing the highest light intensity condition.

These four distinct sets of experimental data undergo optimization algorithm runs to find the optimum parameter combination $\theta_{\text{best-fit}}$. The corresponding fitting results are shown in **Figure S12**, where comparisons of $JV$ characteristics between experimental data and NN output with $\theta_{\text{best-fit}}$ are plotted. The fits were generally poorer when experimental data across different active layer thicknesses were used, compared to fits with light-intensity-dependent data. This observation suggests that, despite being fabricated using the same system, it is difficult to identify common material properties across devices with varying active layer thicknesses. It further implies that variations in the spin-coating process have an impact on morphologies and, thus, on material properties of organic semiconductors[67-72], leading to differences in device performance that are not easily captured by a single set of parameters. The resulting $\theta_{\text{best-fit}}$ values, as suggested by the fitting algorithm, are summarized in **Table 2**. Interestingly, despite analyzing different datasets, some of the extracted parameters exhibit similar values at an approximate level, suggesting that they capture common underlying features of the active layer system. For example, parameters such as $\phi_{\text{an}}$, $f_{\text{ex}}$ and $k_{\text{dir}}$ are kept relatively consistent across the different datasets, while values of $\phi_{\text{cat}}$, $\mu$ and $N_{\text{dt}}$ vary depending on the analyzed dataset.

**Table 2.** Best-fit parameter combination inferred from fitting algorithm for different sets of $JV$ curves. The datasets are categorized into four groups: "thin" and "thick", which comprise light-intensity-dependent $JV$ curves with fixed thicknesses of 83 nm and 247 nm, respectively; and "dim" and "bright", which consist of thickness-dependent $JV$ curves with fixed light intensities of 6 mW/cm² and 91 mW/cm², respectively.

|  | Thin | Thick | Dim | Bright |
| --- | --- | --- | --- | --- |
| Injection barrier anode $\phi_{\text{an}}$ (eV) | 0.26 | 0.24 | 0.28 | 0.31 |
| Injection barrier cathode $\phi_{\text{cat}}$ (eV) | 0.18 | 0 | 0.12 | 0 |
| Exciton dissociation probability $f_{\text{ex}}$ | 0.84 | 0.86 | 0.84 | 0.84 |
| Electron/hole mobility $\mu$ (cm²V$^{-1}$s$^{-1}$) | $5.2 \times 10^{-3}$ | $7.6 \times 10^{-3}$ | $6.1 \times 10^{-3}$ | $1.5 \times 10^{-2}$ |
| Density of deep trap states $N_{\text{dt}}$ (cm$^{-3}$) | $4.0 \times 10^{14}$ | $2.0 \times 10^{15}$ | $9.0 \times 10^{14}$ | $1.1 \times 10^{16}$ |
| Direct recombination coefficient $k_{\text{dir}}$ (cm³s$^{-1}$) | $2.0 \times 10^{-8}$ | $9.5 \times 10^{-9}$ | $1.2 \times 10^{-8}$ | $2.6 \times 10^{-8}$ |

We then utilized these optimal parameters to generate simulated data, varying thickness and light intensities, to predict the device performance, and the results are plotted as solid lines in **Figure 7**. The $\theta_{\text{best-fit}}$ inferred from thickness-dependent data – "dim" and "bright" data – tend to predict the device performance at the light intensity where the dataset is measured but perform poorly at other light intensities. This suggests that datasets at a single light intensity lack information about light-intensity-dependent device performance. Although the $JV$ curves for thin cells were fitted successfully, the predicted efficiencies at higher thickness deviated significantly. Parameter inference run on "thin" dataset overestimated direct recombination coefficient, resulting in poor predictions for thicker devices. In contrast, parameter estimation runs on "thick" dataset yielded better prediction of efficiency over the entire thickness and light intensity range.

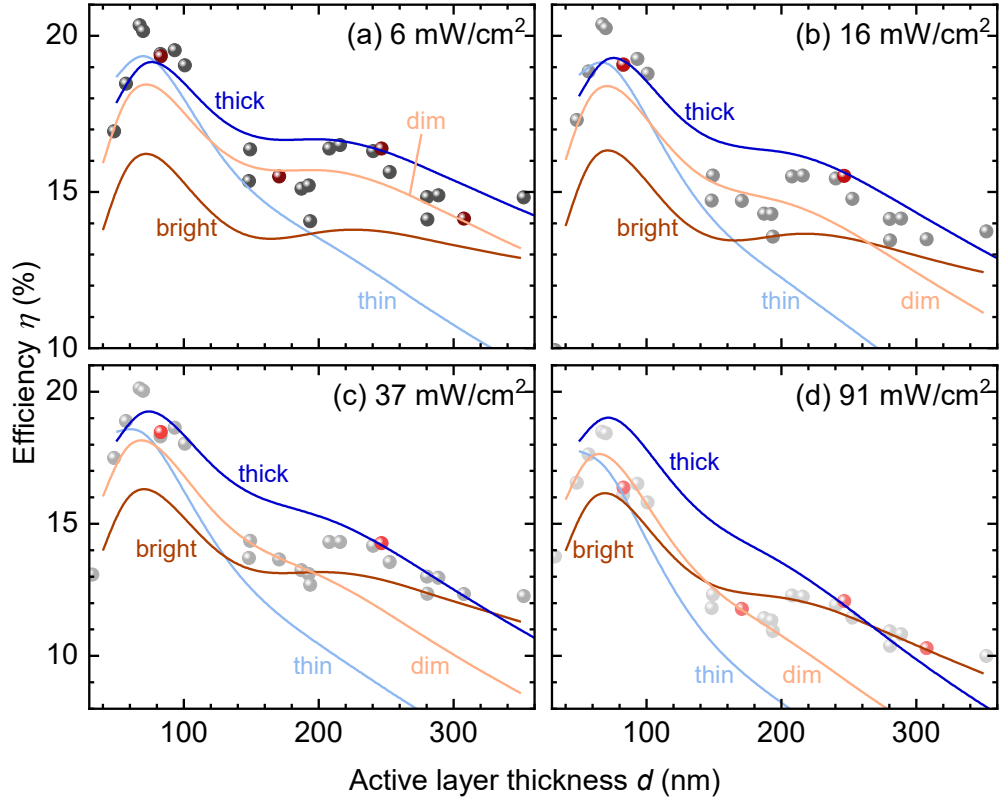

**Figure 7.** Power conversion efficiencies plotted against active layer thickness for light intensity of (a) 6, (b) 16, (c) 37, and (d) 91 mW/cm². The experimental data is represented by dots, with the specific datasets used for Bayesian parameter estimation highlighted in red. The simulated data generated using the best-fit parameter combinations suggested by the fitting algorithm are shown as solid lines. The datasets are categorized into four groups: "thin" and "thick", which comprise light-intensity-dependent *JV* curves with fixed thicknesses of 83 nm and 247 nm, respectively; and "dim" and "bright", which consist of thickness-dependent *JV* curves with fixed light intensities of 6 mW/cm² and 91 mW/cm², respectively. Notably, the parameter estimation run on the "thick" cell dataset yielded the best fit for the experimental data across the entire thickness range. However, the other datasets either failed to converge to a suitable fit for the entire dataset or were able to infer $\theta_{\text{best-fit}}$ but the estimated values do not fit the performance of devices with different thicknesses.

Although the best-fit parameter combinations were identified, none of the runs fully captured the behavior of the system across different thicknesses and light intensities. To examine this further, we compute the probability densities of the six material parameters. While $\theta_{\text{best-fit}}$ is the most probable parameter set determined by the fitting algorithm, the reliability of this combination in accurately predicting outcomes across various thicknesses and light intensities depends on the convergence of the posterior distribution function. In other words, the posterior density functions reflect confidence in the parameter estimates, and poor convergence may indicate that $\theta_{\text{best-fit}}$ is not robust enough to ensure reliable predictions, despite being the algorithm's best fit.

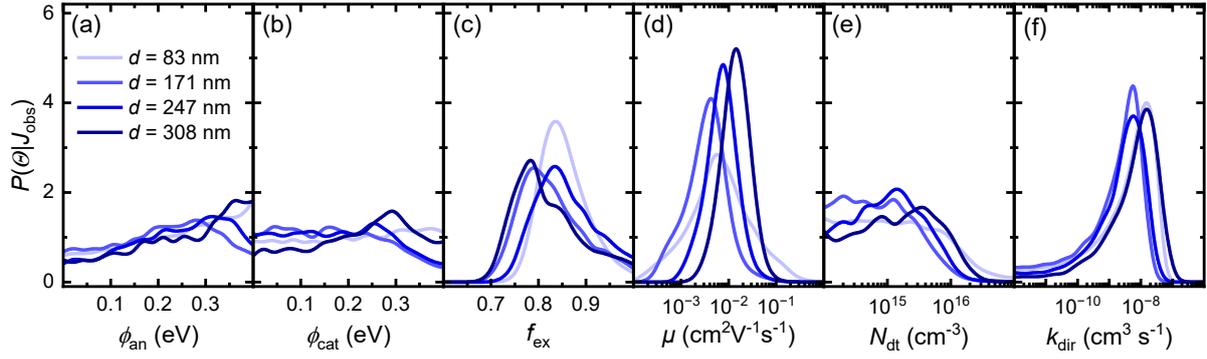

**Figure 8.** Projection of the multivariate posterior probability distribution ($P(\theta|J_{obs})$) onto a one-dimensional parameter space following the observation of the complete set of $JV$ curves. The experimental datasets utilized for parameter estimation consisted of light-intensity-dependent $JV$ curves, each measured at a fixed active-layer thickness of 83, 171, 247, or 308 nm. In general, inferring injection barriers from $JV$ curves is challenging, as evidenced by the uniform distributions of $\phi_{an}$ and $\phi_{cat}$ regardless of the specific datasets. However, certainty in determining defect densities and mobilities increased for the analysis of thicker cells, suggesting that certain types of experiments are more suitable for estimating these parameters.

Therefore, **Figure 8** presents the projected posterior probability distribution of the six material parameters after updating with the complete dataset. As shown in **Figure 7**, the parameter estimation runs on thickness-dependent datasets were unable to converge to a reliable $\theta_{\text{best-fit}}$, likely because of the influence of spin-coating variations on morphologies of the active layer. In the following analysis, we shift our focus to light-intensity-dependent datasets of varied thicknesses with the aim of understanding why parameter estimation runs on thicker cells tend to yield better predictions of device performance. First, the posterior probability distributions of the injection barriers are relatively uniform over the entire range, revealing high uncertainty in estimating $\phi_{an}$ and $\phi_{cat}$, which is consistent with our findings on the synthetic data. In contrast, the probability distributions of $\mu$, $f_{ex}$ and $k_{dir}$ maintain similar shapes across different sample thicknesses, with Gaussian-like distributions and similar peak locations. This observation is consistent with the comparison of the $\theta_{\text{best-fit}}$ values in the previous section. However, the prominence of the peaks varies, indicating differences in the certainty of the parameter estimates. By fitting one of the graphs to a Gaussian function, we can calculate the range of parameter values with high probability. As illustrated in **Figure 9**, the 95% confidence interval of the charge carrier mobility is plotted against the number of observed $JV$ curves, demonstrating that sharper peaks correspond to a more converged, highly probable region. This approach allows us to leverage the full posterior distribution, rather than relying solely on the best-fit parameters.

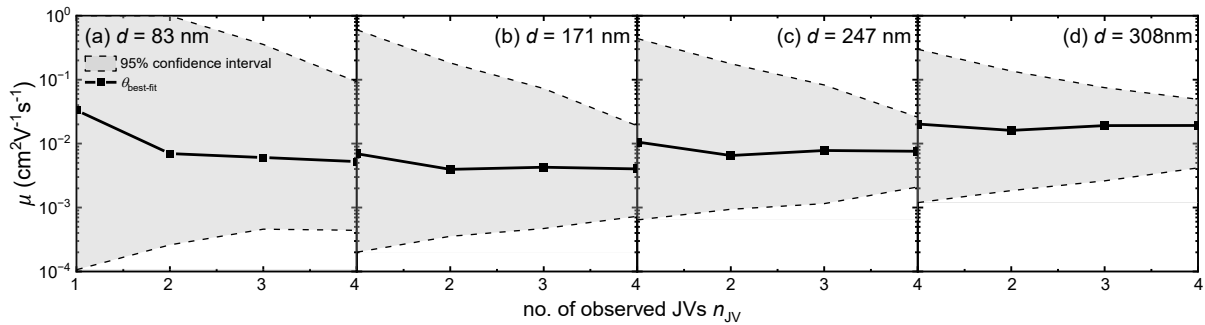

**Figure 9.** The 95% confidence intervals and best-fit parameter values of the charge carrier mobility $\mu$ plotted as a function of the number of observed $JV$ curves, which were obtained by Gaussian fitting of the projected posterior probability distribution shown in Figure 6d. Notably, the analysis runs performed on devices with thicker active layers resulted in reduced uncertainty in estimating the $\mu$, as highlighted by the narrower confidence intervals.

Referring back to **Figure 8**, we observe that the shape of the posterior probability distributions for defect densities varies depending on the thickness of the dataset. In the case of thin cells, the posterior distribution of defect densities is broad and does not exhibit a characteristic peak. However, a distinct peak is revealed in the parameter estimation run on devices with thicknesses of 247 or 308 nm. This can be attributed to space-charge effects: charge accumulation in the active layer due to low mobilities or trapped charge carriers in defect states leading to a redistribution of the internal electric field, hindering charge collection at the contacts. While thin cells with an active layer thickness less than or equal to the width of the space-charge region can maintain a nearly constant electric field, in thicker cells charge accumulation becomes more likely, and, thus, the consequences of space charge become more prominent in the $JV$ characteristics.

To further assess how well the inferred parameters capture the light-intensity-dependent recombination dynamics of the system, we plot the open-circuit voltage $V_{oc}$ as a function of light intensity $\phi$. The slope of the plot provides us the information about the ideality factor $n_{id} = q/kT \times dV_{oc}/d(\ln\phi)$ and thus allows us to identify the dominant recombination mechanism, as shown in **Figure 10**. The experimental data used for parameter estimation is presented as points. Compared with the dashed reference line, both low- and high-thickness cells exhibit ideality factors close to 1, while $n_{id}$ tends to decrease at higher light intensities. The $n_{id} < 1$ may originate from the dominance of surface recombination[49, 73], while this feature could not be adequately captured by adjusting interfacial energy levels alone. Furthermore, to explore the relatively flat posterior distribution of $N_{dt}$ in the lower range of its domain, we varied the defect density while keeping all other parameters fixed at their best-fit values. As shown in **Figure 10a**, for defect densities below $10^{15}$ cm$^{-3}$ (lifetime of 1μs), the ideality factor remains close to 1, making it difficult to distinguish between different defect densities within this range. This may explain the asymmetric shape observed in the posterior probability distribution for $N_{dt}$. Furthermore, the threshold defect density required to maintain an ideality factor near 1 decreases with increasing device thickness, suggesting a stronger influence of defects in thicker devices. This observation implies that certain types of experimental data may be preferable for determining specific material parameters, while $JV$ measurements may not be sufficient for accurately identifying, for example, injection barriers.

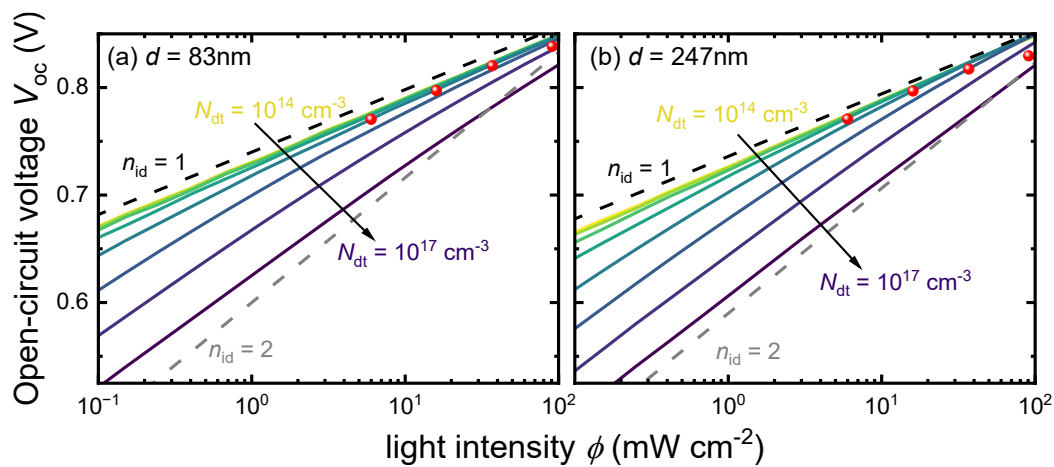

**Figure 10.** Open-circuit voltage plotted against light intensity with varied defect densities $N_{dt}$ in the range between $10^{14}$ and $10^{17}$ cm$^{-3}$ for the thickness of (a) 83 nm and (b) 247 nm. The dashed lines are reference lines indicating the ideality factor $n_{id}$ of 1 or 2. The simulations were conducted with the best-fit values extracted from the respective experimental data represented by red points.

To investigate the interdependencies between material properties of the active layer PBDB-TF-T1:BTP-4F-12, we plot 2-dimensional projected probability density functions, as illustrated in **Figure 11a** and **11c**. The posterior probability results are based on the parameter estimation run on an illumination-dependent experimental dataset with the thickness of 247 nm. We focused on the transport and recombination properties and their correlation among the material parameters. In **Figure 11a**, which shows the joint distribution of $\mu$ and $N_{dt}$, the posterior contours form vertically aligned ellipses, indicating little to no correlation between $\mu$ and $N_{dt}$. In contrast, **Figure 11c**, where $\mu$ and $k_{dir}$ are varied, shows a clear diagonal tilt in the contours, indicating that two parameters are highly correlated to each other. This outcome demonstrates one of the key advantages of Bayesian inference: it not only identifies likely parameter values but also captures dependencies between them, effectively narrowing down the parameter space to plausible regions. For example, while $\mu$ and $N_{dt}$ can vary independently within certain boundaries, $\mu$ and $k_{dir}$ are constrained along a line with a specific slope. To explore how this constrained parameter space affects device performance predictions, we simulated efficiencies across thicknesses using parameter combinations from the reduced space. The results are shown in the solid lines of **Figure 11b** and **11d**, while the experimental results are presented by dots with the red color highlighting the dataset used for this specific run. In **Figure 11b**, defect densities are varied within the 95% confidence interval while other parameters are fixed to best-fit values to simplify the plot. In contrast, both $\mu$ and $k_{dir}$ are varied in **Figure 11d** while maintaining a certain relationship which is extracted from the Gaussian fitting of the plot in **Figure 11c**. Most of the solid lines match well with the red point but provide different predictions for the rest of thickness ranges. This again highlights the benefit of the Bayesian approach: it not only provides a best-fit solution but also a structured way to explore the range of probable combinations. To further narrow down this range of probable parameters and improve confidence in the predictions beyond what is achievable with illumination-dependent $JV$ curves alone, it would be valuable to incorporate additional types of characterization data in future studies.

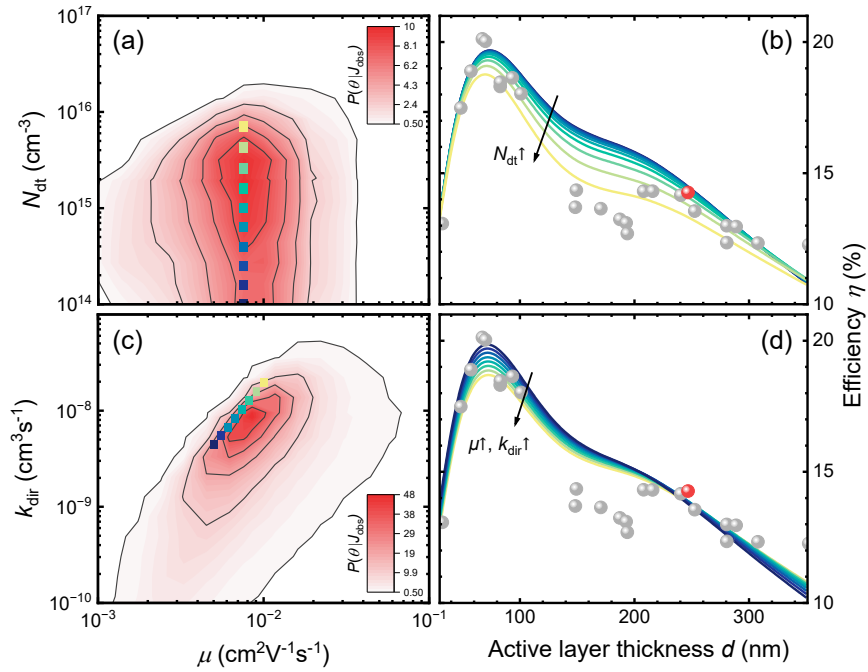

**Figure 11.** (a, c) Contour plots of projected posterior probability distributions onto the 2-dimensional space for the light-intensity-dependent experimental dataset with the active layer thickness of 247 nm. By Gaussian fitting the function, it is found that $\mu$ and $N_{dt}$ are independent of each other, while $\mu$ is highly correlated with $k_{dir}$. (b, d) Power conversion efficiencies plotted against the active layer thickness at a light intensity of 37 mW/cm². The experimental results are shown in dots, where the data used for the parameter estimation run is highlighted with the red color. The solid lines represent the predicted values based on the best-fit parameters $\theta_{best-fit}$ found by the fitting algorithm. In (b), $N_{dt}$ is varied, whereas in (d), $\mu$ and $k_{dir}$ are varied, with the color coding corresponding to the points in the left figure.

## Conclusions

A deep understanding of material systems is essential for accelerating the optimization of organic solar cells. However, the strong correlation between the material properties makes it challenging to isolate individual parameters using experimental data. In the past, numerical simulations, such as drift-diffusion simulations, enabled researchers to reconstruct characterization data using known material parameters and compare the results with experimental data to analyze solar cells fabricated in a lab. Nevertheless, the traditional fitting routine for inferring parameters from characterization data is time-consuming and relies on a deterministic approach that lacks any quantification of confidence in the uniqueness of the resulting fit.

To address these limitations of traditional fitting of numerical models to experimental data in the context of photovoltaics, we applied a methodology from the field of machine learning. We used a neural network as a surrogate model for the device simulation. The role of the neural network is to speed up (by a factor of roughly $10^3$) the device simulation process within a range of predefined parameters. To include information on the confidence in and uniqueness of the resulting fits, we employed a framework of Bayesian Parameter Estimation methods. Therefore, by leveraging machine learning, we efficiently explored the complex parameter space, and the proposed approach provides a more accurate and robust characterization of organic solar cells.

First, we validate our approach by testing it on synthetic data with a known parameter combination. We employed a genetic algorithm to fit the $JV$ curves at various light intensities, and the resulting best-fit parameters were nearly identical to the actual parameters. Furthermore, we performed Bayesian inference runs on the experimental data obtained from organic solar cells with the active layer of PBDB-TF-T1:BTP-4F-12. By analyzing $JV$ curves at different light intensities and for different thicknesses, we were able to infer material properties which comprehensively describe the experimental data to a good approximation. Notably, the values of the extracted injection barrier at anode $\phi_{an}$, exciton dissociation probability $f_{ex}$, and direct recombination coefficient $k_{dir}$ are relatively consistent across different datasets, whereas the $\theta_{\text{best-fit}}$ of injection barrier at cathode $\phi_{cat}$, charge-carrier mobility $\mu$, and deep defect density $N_{dt}$ varied depending on the analyzed dataset. This suggests that while there are common features of the active layer that are independent of the thickness, variations in the spin-coating process can affect the morphologies and material properties of the system. This, in turn, explains why fitting a thickness-dependent dataset is more challenging than the light-intensity-dependent one with a fixed thickness.

Interestingly, we observed that the information gained from the parameter estimation process depends on the dataset used. After performing Bayesian inference runs on light-intensity-dependent datasets with varied thicknesses, we found that the posterior probability distributions of the injection barriers were generally uniform, resulting in low information entropy. In contrast, probability distributions of the mobility and the deep defect density varied depending on the dataset. The significance of the peak in probability density of $\mu$ increased with increasing cell thickness, narrowing down the confidence intervals. On the other hand, a characteristic peak in the probability distribution of $N_{dt}$ emerged only when analyzing thicker cells. This difference in information gain depending on the active layer thickness may arise from more significant space-charge effects in thicker cells, which made the effect more prominent in the $JV$ curve.

While the genetic algorithm can be applied to facile and high-throughput analysis, the suggested probability-based approach enables us to identify the range of highly probable parameters, rather than a single best-fit value. For the cell with the thickness of 247nm, we were able to narrow down the range of mobilities from $10^{-4} - 1$ to $2\times10^{-3} - 3\times10^{-2}$ cm$^2$V$^{-1}$s$^{-1}$ with a confidence level of 95% after observing four light-intensity dependent *JV* curves. Furthermore, the 2D contour plot of joint probabilities allows us to understand the interdependencies of material properties. Our analysis revealed that for the PBDB-TF-T1:BTP-4F-12 layer the mobility and deep defect densities have no correlation while the mobility is highly correlated to direct recombination coefficient. This additional information about material properties enables us to narrow down the probable parameter values. Consequently, we anticipate that our approach of predicting device performance through machine learning-assisted parameter inference will facilitate the optimization of solar cells.

## Acknowledgements

We acknowledge funding by the Helmholtz Association via the POF IV funding as well as via the SolarTap project. We further thank the Ministry of Economic Affairs, Industry, Climate Action and Energy of the State of North Rhine-Westphalia for funding via the ENFA project.

## Data Availability Statement

The data that used in the main text are available in https://doi.org/10.5281/zenodo.15480770. Further information and data are available from the authors upon request.